\documentclass[aps,apl,secnumarabic,twocolumn,showpacs,preprintnumbers,amsmath,amssymb,groupedaddress]{revtex4-1}
\usepackage{graphics}
\usepackage{dcolumn}
\usepackage{bm}
\usepackage[mathscr]{eucal}
\begin{document}

\title{Model for solute diffusion during rapid solidification of binary alloy in semi-infinite volume}

\author{G.L. Buchbinder}
\email[]{glb@omsu.ru}
\author{P. Martaller}
\affiliation{Physics Department, Omsk State University, Peace
Avenue, 55a, 644077 Omsk, Russia\\}

\date{December 8, 2009}

\begin{abstract}
On the basis of local nonequilibrium  approach, the one-dimensional
model of the solute diffusion  during rapid solidification  of the
binary alloy in the semi-infinite  volume  is considered.  Within
the scope  of the model  it is supposed  that  mass transport is
described  by the telegrapher equation.  The basic assumption
concerns   the behavior of the diffusion flux and the solute
concentration at the  interface. Under the condition  that these
quantities  are given by the superposition of the exponential
functions the solutions of the telegrapher equation determining  the
flux and the solute distributions in the melt  have been found.   On
the basis of these solutions  different regimes  of the
solidification in the near surface region  and the behavior of the
partition coefficient have been investigated.  The  concentration
profiles in the solid after complete solidification are analyzed
depending on the model parameters.

\end{abstract}

\pacs{81.10.Aj, 05.70.Fh, 05.70.Ln, 81.30.Fb.}

\maketitle

\section{Introduction}\label{sec:intro}
 In the present  time  the process of rapid
solidification  is a well established method  for production of the
metastable materials and, in particular, supersaturated solid
solutions. The supersaturated metal structures can form during
solidification of binary alloys due
  suppression of the processes  of the solute segregation at the
rapidly moving  solid-liquid  interface \cite{KF92,H94,HGH07}.
Quantitatively this effect can be characterized by the partition
coefficient $k$ defined as the ratio of the solute concentration in
the growing solid to that in the liquid at the sharp interface. The
phenomenon of "solute trapping" by the growing phase, implying, in
particular, that the partition coefficient deviates from its
equilibrium value  $k_e$ and increases towards unity at large growth
rates, has been attracting considerable attention over of several
decades  both from experimental and theoretical points of view
\cite{W80,A82,AK88,AW86,AB94,SA94,EC92,WB93,RB04,C97,AW98,AA99,S95,S96,S97,GS97,G02,GD04,GH06,G07}.

In the rapid solidification experiments very high velocities of the
phase interface can be reached so that the deviations from local
equilibrium in both the bulk of phases and at an interface become
considerable \cite{A82,AK88,BC69,S95,GS97}. For theoretical
description of solute trapping and related phenomena observed during
 rapid solidification  a number models have been proposed
\cite{W80,A82,AK88} in which, in particular, the deviation from
local (chemical) equilibrium at solid-liquid interface is described
by the partition coefficient  $k(V)$ depending on growth velocity
$V$. According to the continuous growth model (CGM) of  Aziz and
Kaplan \cite{AK88} the velocity  depending partition coefficient for
dilute solutions is given in the form
\begin{equation}
    k(V) = \frac{k_e + V/V_{DI}}{1 + V/V_{DI}}\,,\label{Eq1}
\end{equation}
where $V_{DI}$ is the atom diffusive speed at the interface.

The equation (\ref{Eq1}) predicts that the value $k = 1$, that is,
the interface motion without solute partitioning, is only reached at
$V/V_{DI} \gg 1$. Meantime there are  a number of the experimental
works \cite{EC92,BC71,DW60,BC65,M82} in which it has been shown that
the transition to complete solute trapping giving rase  to
diffusionless solidification  occurs at substantially finite  values
of $V/V_{DI}$. This circumstance is automatically taken into account
 within the scope  of the local nonequilibrium approach developed in the works \cite{S95,S96,S97,GS97,G02,GD04,GH06,G07}.
At the high growth velocities  the deviation from local equilibrium
can be essential not only at the interface  but in the bulk of the
liquid phase as well. The extension of the expression (\ref{Eq1})
for this case introduced in \cite{S96} has the form
\begin{equation}\label{Eq2}
k(V) = \left\{\begin{array}{ll}
\displaystyle{\frac{(1 - V^2/V^2_{D})k_e + V/V_{DI}}{1 - V^2/V^2_{D} + V/V_{DI}}}\,,&   V < V_{D}\\
1\,,& V \geqslant V_{D}\,.
\end{array}\right.
\end{equation}
The expression (\ref{Eq2}) takes naturally into account the fact
that when the growth velocity $V$  exceeds  the velocity of the
propagation of concentration disturbances in the liquid phase $V_D$
the solute transfer  in a melt is absent and the transition to the
diffusionless solidification occurs at the finite velocity  $V =
V_D$.  The extensions of  expressions (\ref{Eq1}) and (\ref{Eq2}) to
the application to the concentrated alloys are given in
\cite{AK88,G07}.

It  should be noted that the currently existing  theoretical models
consider the phase interface far from the surface of a melt  that is
in fact an infinite system. The presence of the surface leads  to
the additional peculiarities.  If the matter flux  from the surface
is absent, complete solute trapping by the growth phase must occur
in the near surface layers. As a result, the partition coefficient
must tend to unity when the interface approaches  the free surface
even with small velocities  $V$ and can vary in a complex enough way
in the near surface region except for the case of $V \geqslant V_D$.

The behavior of the key quantities, characterizing a solidification
process,  at the moving interface  specifies eventually the micro -
and macrostructure of the resulting solid  and therefore is a
subject of the central interest.

Rapid solidification of dilute Ni(Zr) alloy after pulsed laser
irradiation has been studied  in the work of Arnold et. al.
\cite{AA99}. A simulation  of the distribution of Zr after complete
solidification using the CGM has shown that the simulated
concentration profile reproduces  the experimental data well enough
in the deep regions. However, there is a significant discrepancy in
data  in the near surface region. It has been assumed in \cite{AA99}
that an initially planar interface becomes unstable in the near
surface region  and degenerates into a cellular structure although
the authors have not been able to verify this experimentally. On the
other hand, a one-dimensional simulation using a planar interface
and the partition coefficient increasing with time  in near surface
region as
\begin{equation}\label{Eq3}
    k(t) \thicksim e^{t/\tau}\,,
\end{equation}
where $\tau$ is a fitting parameter, has  given good enough
agreement with the data  in near surface too.

The relationship  (\ref{Eq3}) enables one to make some assumption
about the behavior  of the solute concentration at the interface.
Indeed, according to the relationship (\ref{Eq3}) the dependence  of
the partition coefficient on the position of the interface $x = -
Vt$, moving with a constant velocity  in parallel to the free
surface  (fixed at $x = 0$) has  the form
\begin{equation}\label{Eq4}
   k(x) \thicksim e^{ - x/l},
\end{equation}
with $l = \tau V$. Because $k$ is defined as the ratio of the solute
concentrations  in different phases, the relationship (\ref{Eq4})
means in the simplest case  that the solute concentrations taken at
the interface in  each phase can vary exponentially as well but with
different exponents. In a more general case these quantities can be
defined as the superposition of the terms of the form (\ref{Eq4}).

In the present work the one-dimensional model of the solute
diffusion during rapid solidification of the dilute binary alloy  in
the semi-infinite volume is considered. The interface velocity  is
presumed  to be high enough to provide stability of the planar front
of the growth of the  solid phase \cite{KT90,GD04}.   It is also
presumed that the motion of the interface is so fast  that local
equilibrium  in the bulk  of the liquid phase is absent  and the
solute transport occurs under local nonequilibrium conditions
\cite{S95,S96,S97,GS97,G02,GD04,GH06,G07}. Such conditions are
achieved by high undercooling of the melt or during
recrystallization after laser irradiation of the solid. The main
purpose of this work is the development of the model determining the
behavior of the solute concentration and the diffusion flux both at
the fast moving interface  and in the bulk of the phases and the
determination of the inhomogeneous partition coefficient.

The work is organized as follows.  In Sec. \ref{sec:mod} the general
equations describing  the local nonequilibrium transport are
formulated and the model defining the behavior  at the interface  of
the key quantities of interest is given.  The solutions of the
transport equations for semi-infinite volume have been obtained by
and the Riemann method for the hyperbolic differential equations.
  These  solutions define the diffusion flux in the
 liquid phase and the solute concentration  in
 both phases. The  inhomogeneous partition coefficient is
derived. The discussion and the conclusion are respectively given in
Sec. \ref{sec:dis} and Sec. \ref{sec:con}. The Riemann method in
detail  and its application to the present problem are given in
Appendixes.
\section{The Model}\label{sec:mod}
As it has been noted above the interface  velocity can reach high
values. This  takes place, for example, at the high undercooling  of
the melt or  during solidification after the laser irradiation  of
the surface  of an alloy.  When  the interface velocity $V$ is equal
or more than the velocity of the propagation of the concentration
disturbances in the liquid phase $V_D$, the diffusion field  in the
liquid  can significantly deviate from local equilibrium
\cite{S95,GS97}. In this case the solute diffusion flux is no longer
defined by the classical Fick's law relating  the diffusion flux  to
the gradient  of a concentration, and  it should be considered as an
independent variable as well as the solute concentration. According
to extended irreversible  thermodynamics \cite{JCL96} the simplest
generalization of the Fick's law taking into account the relaxation
to local equilibrium  in the diffusion field is given  by the
  Maxwell-Cattaneo equation which  one writes down in the
one-dimensional form as
\begin{equation}\label{Eq5}
   J_L + \tau\frac{\partial J_L}{\partial t} = - D_L \frac{\partial C_L}{\partial
   x}\,,
\end{equation}
where index $L$ relates  to the liquid phase, $J_L$ is the solute
diffusion flux, $C_L$ is the solute concentration, $\tau$ is the
time of relaxation of the diffusion flux to its local equilibrium
value defined by the Fick's law and $D_L$ is the diffusion
coefficient.

 Eq.~(\ref{Eq5}) in combination with the conservation law
\begin{equation}\label{Eq6}
   \frac{\partial C_L}{\partial t} = -  \frac{\partial J_L}{\partial
   x}\,,
\end{equation}
gives rise to the hyperbolic transport equations
\begin{eqnarray}
  \tau\frac{\partial^2 C_L}{\partial t^2} + \frac{\partial C_L}{\partial t} &=& D_L\frac{\partial^2 C_L}{\partial x^2} \label{Eq7}\\
 \tau\frac{\partial^2 J_L}{\partial t^2} + \frac{\partial J_L}{\partial t} &=& D_L\frac{\partial^2 J_L}{\partial x^2}
 \label{Eq8}\,.
\end{eqnarray}
The equation of the type (\ref{Eq7}) and (\ref{Eq8}) is known as the
telegrapher  equation that combines the properties  both of the wave
equation and the diffusion one. At the times of the order  $\tau$ it
predicts the finite velocity of the propagation of concentration
disturbances  $V_D = (D_L/\tau )^{1/2}$ in contrast  to the
diffusion equation for which $V_D = \infty$ at all time scales.

To describe the mass transport  during the solidification process we
consider a binary melt initially occupying half-space  $x \geqslant
0$. The planar front of solidification  forms  in the infinitely
removed region at $t = - \infty$ and isothermally moves with the
constant average  velocity $V$ to the surface  of the system, fixed
at $x = 0$, along the trajectory $x + Vt = 0$ in parallel  to the
free surface.  At an arbitrary moment of time  the region occupied
by the melt is given  by the inequality  $0 \leqslant x \leqslant -
Vt$ $(t \leqslant 0)$. Therefore in the plane $(x, t)$ the liquid
phase occupies the region $ x + Vt \leqslant 0$, $x\geqslant 0$,
$t\leqslant 0$. At the interface representing the surface of a
discontinuity  the conservation law of mass holds \cite{LL87} that
in accepted notations has the form
\begin{equation}\label{Eq9}
 [J_L + VC_L]_{x + Vt = 0} = [J_S + VC_S]_{x + Vt = 0}\,,
\end{equation}
where index S  relates to the solid.  Taking into account a small
mobility of the solute in the solid by comparison with its  mobility
in the liquid phase one neglects, as usual,  by diffusion in the
solids  and writes down Eq.~(\ref{Eq9}) as
\begin{equation}\label{Eq10}
   V(C_L - C_S)|_{x + Vt = 0} = - J_L|_{x + Vt = 0}
\end{equation}
Now we consider the diffusion flux in more detail. Introducing
dimensionless variables $t/\tau$, $x/\tau V_D$ in Eq.~(\ref{Eq9})
one obtains
\begin{equation}\label{Eq11}
   \frac{\partial^2 J}{\partial t^2} + \frac{\partial J}{\partial t} = \frac{\partial^2 J}{\partial x^2}
 \,,
\end{equation}
where   the former notations  $(x, t)$  have been used for new
variables and $J/V_D$ is the dimensionless diffusion flux.  The
boundary condition (\ref{Eq10}) in the dimensionless form is written
as
\begin{equation}\label{Eq12}
    \alpha (C_L - C_S)|_{x + \alpha t = 0} = - J|_{x + \alpha t = 0}\,
\end{equation}
where  the dimensionless parameter $\alpha = V/V_D$  characterizes
the extent  of the deviation  of the system  from local equilibrium.
In addition, at the surface  the equality  should be fulfilled
\begin{equation}\label{Eq13}
   J(xt)|_{x=0} = 0\hspace{1cm}(t \leqslant 0)\,,
\end{equation}
expressing the condition of the absence  of the flux through the
surface. At last, the solution of  Eq.~(\ref{Eq11}) is sought in the
region
$X \equiv x + \alpha t \leqslant 0$, $x\geqslant 0$, $t\leqslant 0$
occupied by the liquid phase while the solid occupies the region $X
\geqslant 0$.

It is physically apparent that at $\alpha \geqslant 1$, that is $V
\geqslant V_D$,  the presence of the surface  is not of considerable
importance. Interface  moves with the velocity equal to or exceeding
the velocity of the propagation of the concentration disturbances
and the solute distribution  in the liquid phase remains
homogeneous. The solution of  Eq.~(\ref{Eq11}) satisfying  this
condition  and compatible with the equality (\ref{Eq13}) is
\cite{S95,GS97}
\begin{equation}\label{Eq15}
 J = 0,\hspace{0.5cm} C_L = C_S = \mbox{const}\hspace{0.5cm}(\alpha \geqslant
 1),
\end{equation}
which corresponds to the complete solute trapping by the growth
phase.

Now we consider the case of $\alpha < 1$. Suppose that at the moving
interface residing in an arbitrary point $x$ at the moment $t = -
x/\alpha$ the flux $J$ and its the time derivative $\partial J
/\partial t$ are  known
\begin{eqnarray}
 J(x t)|_{t = - x/\alpha}  &=& j_0(x)\,, \label{Eq16}\\
  \displaystyle\frac{\partial J(x t)}{\partial t}|_{t = - x/\alpha}  &=& j_1(x)\,,\label{Eq17}
\end{eqnarray}
where the function  $j_0(x)$ and $j_1(x)$  will be specified
further. Eqs.~(\ref{Eq16}) and ~(\ref{Eq17}) determinate the
"initial" conditions  that are given at the straight line $x +
\alpha t =0$ defining the trajectory  of the interface  rather than
at $t = 0$.

If the functions $j_0(x)$ and $j_1(x)$ are known the solution of
Eq.~(\ref{Eq11}) satisfying the conditions (\ref{Eq16}) and
(\ref{Eq17}) in the region  $X \leqslant 0$ at $\alpha < 1$ can be
found  by the Riemann method \cite{TS04} (for details see Appendix
A) and has the form
\begin{widetext}
\begin{eqnarray}
 J(xt)& =& \frac{1}{2}\biggl \{\varphi \Bigl (- \alpha\frac{x + t}{1 - \alpha}\Bigr)\exp\Bigl[\frac{X}{2(1 - \alpha)}\Bigr]  +
 \varphi \Bigl( \alpha\frac{x - t}{1 + \alpha}\Bigr)\exp \Bigl [-\frac{X}{2(1 + \alpha)}\Big]\biggr\} - \nonumber \\
  & - &\frac{1}{2}e^{- t/2}\int\limits_{\displaystyle{-\frac{\alpha (x + t)}{1 - \alpha}}}^{\displaystyle{\frac{\alpha (x - t)}{1 + \alpha}}}
  \,dx_1 \psi (x_1)e^{- x_1/2\alpha}J_0 \Bigl (\frac{1}{2}\sqrt{(x - x_1)^2 - (t + x_1/\alpha)^2}  \Bigr ) + \nonumber\\
  & + & \frac{X}{4\alpha}e^{- t/2}\int\limits_{\displaystyle{-\frac{\alpha (x + t)}{1 - \alpha}}}^{\displaystyle{\frac{\alpha (x - t)}{1 + \alpha}}}
  \,dx_1 \varphi (x_1)e^{- x_1/2\alpha}\frac{J_0' \Bigl (\frac{1}{2}\sqrt{(x - x_1)^2 - (t + x_1/\alpha)^2}  \Bigr
  )}{\sqrt{(x - x_1)^2 - (t + x_1/\alpha)^2}}\,,\label{Eq18}
 \end{eqnarray}
   \end{widetext}
where
\begin{eqnarray}
  \varphi (x) &=& j_0(x) \label{Eq19}\\
  \psi (x) &=& \frac{1}{2}j_0(x) - \frac{1}{\alpha }j_0'(x) - \frac{1 - \alpha ^2}{\alpha
  ^2}j_1(x)\label{Eq20}
\end{eqnarray}
and $J_0(x)$ is the Bessel function of zero order. At arbitrary
$j_0(x)$ and $j_1(x)$ (or $\varphi (x)$ and $\psi (x)$) the
expression (\ref{Eq18}), in general, does not satisfy   the boundary
condition (\ref{Eq13}).

Now we consider  the model  within the scope  of which  (as it has
been discussed in Introduction) {\it all the quantities given at the
phase interface}  are represented  by the linear  combinations of
the exponential functions. In particular, let $\varphi (x)$ and
$\psi (x)$ be given  by the expansions
\begin{eqnarray}
  \varphi (x)  &=&  A_0 + A_1e^{- \gamma_1 x/2} + A_2e^{- \gamma_2 x/2} + \cdots\,,\label{Eq21}\\
 \psi(x) &=& B_0 + B_1e^{- \gamma_1 x/2} + B_2e^{- \gamma_2
 x/2} + \cdots\,,\label{Eq22}
 \end{eqnarray}
 where constants $\gamma_n \geqslant 0$, $A_n$ and $B_n$ will be specified in
 what follows. After the substitution of  Eqs.~(\ref{Eq21})
 and (\ref{Eq22}) in  Eq.~(\ref{Eq18}) and the calculation of
 the integrals (details see in Appendix B), we obtain
\begin{widetext}
\begin{equation}
   J_n(x t) = \sum\limits_{n\geqslant 0}
   e^{- \gamma_n x/2}\Bigl\{ A_n^{(-)}\exp\Bigl[\frac{\gamma_n^{(+)}X}{2(1 - \alpha^2)}\Bigr
   ]
   +  A_n^{(+)}\exp\Bigl[\frac{\gamma_n^{(-)}X}{2(1 - \alpha^2)}\Bigr ]
\Bigr\} \label{Eq24}
\end{equation}
\end{widetext}
and the notations have been introduced
 \begin{eqnarray}
   \gamma_n^{(\pm)} &=&  \gamma_n + \alpha \pm \sqrt{\alpha^2\gamma_n^2 + 2\alpha\gamma_n + \alpha^2} \geqslant 0\,; \label{Eq25}\\
   A_n^{(\pm)}&=& \frac{A_n}{2} \pm  B_n\frac{\delta_n}{\nu_n} \,; \label{Eq26} \\
   \delta_n &=& \frac{\alpha}{1 + \alpha\gamma_n}\,; \label{Eq27}  \\
    \nu_n&=&\sqrt{1 - \frac{\delta_n ^2}{\alpha^2}(1 - \alpha^2)}\,. \label{Eq28}
 \end{eqnarray}
 As it is seen from Eqs.~(\ref{Eq21}) and (\ref{Eq22}) $A_0$ and
 $B_0$ determinate the behavior of $\varphi$ and $\psi$ (or $j_0$ and $j_1$
 ) far from the system surface. Let us determinate the rest of the parameters
 $\gamma_n$, $A_n$ and $B_n$ ($n\geqslant 1$) in such a way  as to
 satisfy the balance condition (\ref{Eq12}) and the boundary
 condition at the free surface (\ref{Eq13}).

\renewcommand{\thesubsection}{\Alph{subsection}}
\subsection{\label{sec:par}The determination of the parameters}
Now consider the boundary condition (\ref{Eq13}). Taking into
account  that $\gamma _0 = 0$, $\delta_0 =\alpha$,  $\nu_0 =
\alpha$, $\gamma^{(\pm)}_0 = \alpha \pm\alpha$ and using
Eq.~(\ref{Eq24}), we have for an arbitrary   $t < 0$
\begin{widetext}
\begin{eqnarray}
  J(x, t)|_{x = 0} &=& \Bigl (\frac{A_0}{2} - B_0\Bigr )\exp{\frac{2\alpha^2t}{2(1 - \alpha^2)}} +  \Bigl (\frac{A_0}{2} + B_0\Bigr ) +\nonumber\\
  &=&\Bigl (\frac{A_1}{2} - B_1\frac{\delta_1}{\nu_1}\Bigr )\exp{\frac{\gamma ^{(+)}_1\alpha t }{2(1 - \alpha^2)}} + \Bigl (\frac{A_1}{2} + B_1\frac{\delta_1}{\nu_1}\Bigr )\exp{\frac{\gamma ^{(-)}_1\alpha t }{2(1 - \alpha^2)}} + \nonumber \\
   &=& \Bigl (\frac{A_2}{2} - B_2\frac{\delta_2}{\nu_2}\Bigr )\exp{\frac{\gamma ^{(+)}_2\alpha t }{2(1 - \alpha^2)}} +
   \Bigl (\frac{A_2}{2} + B_2\frac{\delta_2}{\nu_2}\Bigr )\exp{\frac{\gamma ^{(-)}_2\alpha t }{2(1 - \alpha^2)}} + \cdots = 0\,.\label{Eq29}
\end{eqnarray}
\end{widetext}

If all the powers of the exponentials are different then $J(0, t) =
0$ can be only at $A_n = B_n = 0$. However if each exponential
function will appear
 in Eq.~(\ref{Eq29}) at least twice then this can lead to nonzero $A_n$
 and $B_n$. Bearing in mind this circumstance we determinate
 $\gamma_n$ so that the equalities are held
\begin{equation}\label{Eq30}
    \gamma^{(-)}_n = \gamma^{(+)}_{n - 1 }\hspace{2cm}n = 1, 2, 3,\ldots\;,
\end{equation}
in which $\gamma^{(+)}_{n - 1 }$ (and respectively $\gamma_{n - 1}$)
are considered to be known \cite{COM}. Taking into account  the
notation (\ref{Eq25}) and resolving the Eq.~(\ref{Eq30}) in relation
to $\gamma_n$, one obtains
\begin{equation}\label{Eq31}
    (\gamma_n)_{12} = \frac{\gamma^{(+)}_{n - 1 } \pm \sqrt{\alpha\gamma^{(+)}_{n - 1 }[2(1 - \alpha^2) + \alpha\gamma^{(+)}_{n - 1
    }]}}{(1 - \alpha^2)} \,\:.
\end{equation}
\begin{table*}
\caption{\label{tab:table1}The parameters $\gamma_n$ appearing in
Eqs.~(\ref{Eq21}) and (\ref{Eq22}).}
\begin{ruledtabular}
\begin{tabular}{cccccc}
n&0&1&2&3&4\\
\hline
 $\gamma_n$ &$0$ &${\displaystyle\frac{4\alpha}{1 - \alpha^2}}$ &${\displaystyle\frac{4\alpha(3 + \alpha^2)}{(1 - \alpha^2)^2}}$
 &${\displaystyle\frac{8\alpha (3 + \alpha^2)(1 + \alpha^2)}{(1 - \alpha^2)^3}}$
 &${\displaystyle\frac{8\alpha(1 + \alpha^2)(\alpha^4 +10\alpha^2 + 5)}{(1 - \alpha^2)^4}}$ \\

 $\gamma_n^{(+)}$ &$2\alpha$  &${\displaystyle\frac{8\alpha}{1 - \alpha^2}}$
 &${\displaystyle\frac{2\alpha(3 + \alpha^2)^2}{(1 -
 \alpha^2)^2}}$&${\displaystyle\frac{32\alpha(1 + \alpha^2)^2}{(1 - \alpha^2)^3}}$
 &$  {\displaystyle\frac{2\alpha (\alpha^4 +10\alpha^2 + 5)^2}{(1 - \alpha^2)^4}}  $\\

$\gamma_n^{(-)}$ &0 &$2\alpha$ &${\displaystyle\frac{8\alpha}{1 -
\alpha^2}}$ &  ${\displaystyle\frac{2\alpha(3 + \alpha^2)^2}{(1 -
\alpha^2)^2}}$
&${\displaystyle\frac{32\alpha(1 + \alpha^2)^2}{(1 - \alpha^2)^3}}$\\
\end{tabular}
\end{ruledtabular}
\end{table*}
At $n = 1$ and $ \gamma^{(+)}_0 = 2\alpha$  Eq.~(\ref{Eq31}) gives
\[\gamma_1 =  \frac{4\alpha}{1 - \alpha^2} .\]
The second value $\gamma_1 = 0$ is the extraneous root of the
Eq.~(\ref{Eq31}) at $n = 1$. After the determination of $\gamma_1$
the values $\gamma^{(\pm)}_1$ appearing in Eq.~(\ref{Eq24}) can be
found from Eq.~(\ref{Eq25}). Along similar a line one can obtain the
values $\gamma_n$, $\gamma^{(\pm)}_n$ for $n >1$. In
Table~\ref{tab:table1} these values are given for $n \leq 4$. As it
is seen from the table $\gamma_n \sim (1 - \alpha^2)^{- n}$,
 $\gamma_n^{(+)} \sim (1 - \alpha^2)^{- n}$ è $\gamma_n ^{(-)}\sim (1 - \alpha^2)^{- n +
 1}$.  The case of an arbitrary $n$ is easily  proved by induction
 using Eq.~(\ref{Eq31}). Then considering the inequality $1 - \alpha^2 \ll
 1$ we neglect by the exponentially small terms in the sum (\ref{Eq24})
 and restrict ourselves by the  terms with $n \leqslant 3$ only. To do this it will
 suffice to put
\begin{equation}\label{Eq32}
    A^{(-)}_3 = \frac{A_3}{2} - B_3\frac{\delta_3}{\nu_3} = 0,
    \hspace{0.5cm} A_n = B_n = 0,\: n \geqslant 4,
\end{equation}
that corresponds to the first four terms of the expansions
(\ref{Eq21}) and (\ref{Eq22}).

Now  let us define the rest of the nonzero constants $A_n$ and $B_n$
so that the condition (\ref{Eq29}) holds. As a result, the constants
$B_n$ are completely eliminated and we have finally
\begin{eqnarray}
   J(x,t) = A_0(1 - e^{- \gamma_1x/2})\exp\frac{\alpha X}{1 - \alpha^2} \phantom{(A_2 + A_3 + A_3)}&&\nonumber \\
 + (A_2 + A_3)(e^{- \gamma_2x/2} - e^{- \gamma_1x/2})\exp\frac{4\alpha X}{(1 - \alpha^2)^2}\phantom{(A_2 }&&
    \nonumber\\
     + A_3(e^{- \gamma_3x/2} - e^{- \gamma_2x/2})\exp\frac{\alpha (3 + \alpha^2)^2 X}{(1 - \alpha^2)^3}\,,\phantom{(A_2 }&&\label{Eq33}
 \end{eqnarray}
The simple but rather cumbersome calculations show that the flux
defined by Eq.~(\ref{Eq33}) satisfies  the initial conditions
(\ref{Eq16}) and (\ref{Eq17}).

It should be noted that the expression (\ref{Eq33}) can be
represented in the alternative form explicitly demonstrating the
presence of the wave component in the mechanism of the solute
transport
\begin{widetext}
\begin{eqnarray}
  J(x, t) &=& A_0\Bigl[ \exp\frac{\alpha X}{1 - \alpha^2} -  \exp \Bigr (-\frac{\alpha \tilde{X}}{1 - \alpha^2}\Bigr )  \Bigr ] +\nonumber \\
  &+& (A_2 + A_3)\Biggl\{ \exp\Bigl [\frac{\gamma_1x}{2} - \frac{ 4\alpha \tilde{X}}{(1 - \alpha^2)^2}\Bigr ] -  \exp\Bigl [-  \frac{\gamma_1x}{2} +\frac{ 4\alpha X}{(1 - \alpha^2)^2}\Bigr ] \Biggr \} + \nonumber   \\
  &+&  A_3\Biggl\{ \exp{\Bigl [\frac{\gamma_2x}{2} - \frac{ \alpha (3 + \alpha^2)^2 \tilde{X}}{(1 - \alpha^2)^3}\Bigr ]} -
   \exp\Bigl [-  \frac{\gamma_2x}{2} +\frac{ \alpha (3 + \alpha^2)^2 X}{(1 - \alpha^2)^3}\Bigr ]  \Biggr
   \}\,,\label{Eq35}
\end{eqnarray}
\end{widetext}
where
\[ X = x + \alpha t \leqslant0; \hspace{0.5cm} \tilde{X} = x - \alpha t
\geqslant 0;\hspace{0.5cm}(x \geqslant0, t \leqslant0)\,.\] As it is
seen from Eq.~(\ref{Eq35}) the terms containing $\tilde{X}$ can be
considered as the concentration waves reflected from the surface and
propagating   to the interface.

Except for the condition \[A_0 + A_1 +A_2 +A_3  =  j_0(0) = J(0, 0)
= 0\,,\] following from Eqs.~(\ref{Eq19}) and  (\ref{Eq21}), the
constants $A_n$  appearing  in Eqs. (\ref{Eq33}) and (\ref{Eq35}),
remain up to now  arbitrary and must be defined from other
conditions that we shall consider in the next section.
\begin{figure*}
\includegraphics{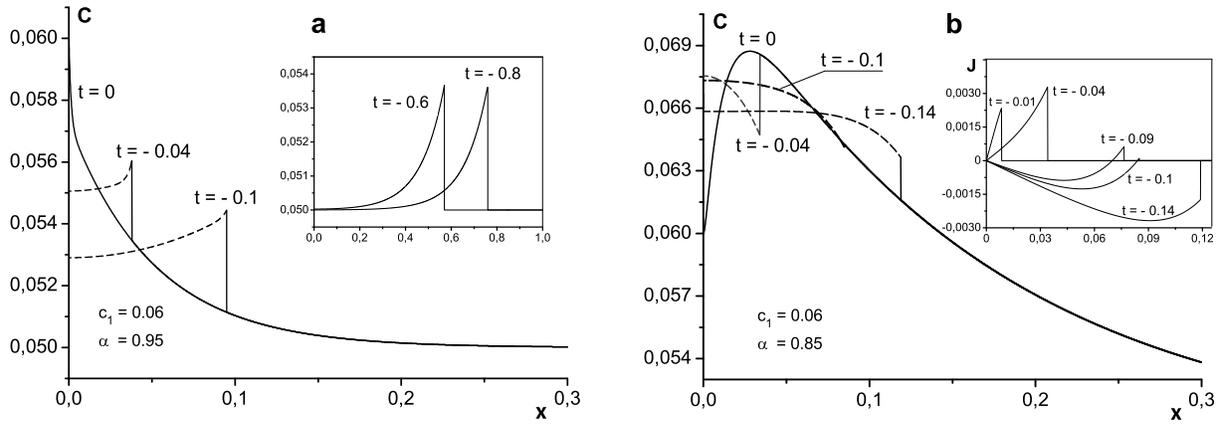}
\caption{\label{fig1} The concentration profiles for different
values  of $\alpha = V/V_D$ and the surface solute concentration
$c_1 = C_S(0)$ in the solid. The material constants are $V_D = 25
(m/s)$, $V_{DI} = 20 (m/s)$, $k_e = 0.1$, $c_0 = 0.05$. The solid
lines are the solute profiles in the solid phase, the dash lines are
the solute profiles in the liquid one. a) $\alpha = 0.95$, $c_1 =
0.06$. In the inset the concentration profiles are shown when the
interface is far from the surface. In  both phases the curves are
depicted by the solid lines.\\  b) $\alpha = 0.85$, $c_1 = 0.06$. In
the inset the solute flux in the liquid phase is shown.}
\end{figure*}
\renewcommand{\thesubsection}{\Alph{subsection}}
\subsection{\label{sec:conc}The solute concentration}
The solute concentration $C_L$ in the liquid phase can be found  in
the same way as the flux has been defined. Let
\begin{eqnarray}
 C_L(x t)|_{t = - x/\alpha}  &=& c_L(x)\,, \nonumber\\
 \displaystyle\frac{\partial C_L(x t)}{\partial t}|_{t = - x/\alpha}  &=& c_{L1}(x)\,,\nonumber
\end{eqnarray}
and  at the interface  the relationships of the type  of the
expansions  (\ref{Eq21}) and (\ref{Eq22})  take place
\begin{eqnarray}
  \varphi_c (x)  & =&  a_0^{(L)} + a_1^{(L)}e^{- \gamma_1 x/2} + a_2^{(L)}e^{- \gamma_2 x/2} + \cdots\,,\nonumber\\
 \psi_c(x) &=& b_0 + b_1e^{- \gamma_1 x/2} + b_2e^{- \gamma_2
 x/2} + \cdots\,,\label{Eq36}
 \end{eqnarray}
where $\varphi_c (x)$ and $\psi_c(x)$ are related to $c_L(x)$ and
$c_{L1}(x)$  by the equalities analogous to  the Eqs.~(\ref{Eq19})
and (\ref{Eq20}). Then the solution of Eq.~(\ref{Eq7}) (the latter
is written down in the dimensionless form) has the form similar to
 the solution (\ref{Eq35})  where  $A_n$ and $B_n$ must be
replaced by $a_n^{L}$ and $b_n$. For determination of this
parameters the mass conservation law can be used. Substituting the
expression (\ref{Eq35})  and the corresponding expression for the
solute concentration $C_L(x, t)$  into Eq.~(\ref{Eq6}) and equating
the coefficients at the linear independent  functions one can write
down  $a_n^{(L)}$ and $b_n$ in terms $A_n$. As a result, using the
found values for $\gamma_n$ one has  within the approximation as for
Eq.~(\ref{Eq35})
\begin{widetext}
\begin{eqnarray}
& & C_L(x, t)=  c_0 - \frac{A_0}{\alpha}\Bigl[ \exp\frac{\alpha X}{1 - \alpha^2} + \exp\Bigl (- \frac{\alpha \tilde{X}}{1 - \alpha^2}\Bigr )  \Bigr ]  \nonumber \\
 & + & \frac{(1 + \alpha^2)}{2\alpha}(A_2 + A_3)\Biggl\{ \exp\Bigl
[\frac{\gamma_1x}{2} - \frac{ 4\alpha \tilde{X}}{(1 -
\alpha^2)^2}\Bigr ] +
  \exp\Bigl [- \frac{ \gamma_1x}{2} +\frac{ 4\alpha X}{(1 - \alpha^2)^2}\Bigr ] \Biggr \} \nonumber   \\
&  + & \frac{(1 + 3\alpha^2)A_3}{\alpha(3 + \alpha^2)}\Biggl\{
\exp{\Bigl [\frac{\gamma_2x}{2} - \frac{ \alpha (3 + \alpha^2)^2
\tilde{X}}{(1 - \alpha^2)^3}\Bigr ]} +
   \exp\Bigl [-  \frac{\gamma_2x}{2} +\frac{ \alpha (3 + \alpha^2)^2 X}{(1 - \alpha^2)^3} \Bigr ]  \Biggr
   \}\,,\label{Eq37}
\end{eqnarray}
\end{widetext}
where $c_0$ is the initial solute concentration in the melt
\[c_0 =
\lim_{t \rightarrow - \infty}C_L(x, t)\,.\]
\begin{figure}
\includegraphics[10,25][570,200]{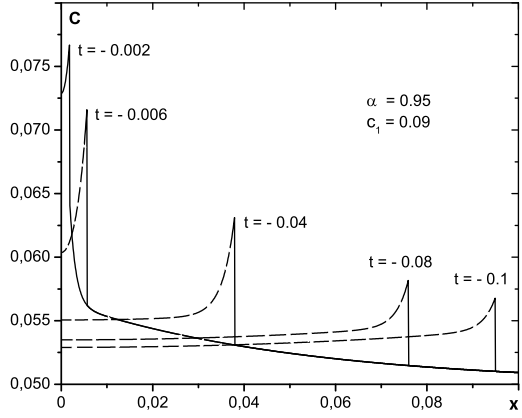}
\caption{\label{fig2} The concentration profiles  for $\alpha =
0.95$, $c_1 = 0.09$, $c_0 = 0.05$, $V_D = 25 (m/s)$, $V_{DI} = 20
(m/s)$, $k_e = 0.1$. The solid lines are the solute profiles in the
solid phase, the dash lines are the solute profiles in the liquid
one. }
\end{figure}
At last, taking into account that at the interface
\[c_L(x) = C_L(x, t)|_{X = 0} = \varphi_c(x)
\,,\] one obtains  for the coefficients of the expansion
(\ref{Eq36})
\newpage
\begin{eqnarray}
  a_0^{(L)} &\equiv& c_L^{\infty} = c_0 -\frac{A_0}{\alpha}  \,, \nonumber\\
  a_1^{(L)}&=& \frac{1 + \alpha^2}{2\alpha}(A_2 + A_3) - \frac{A_0}{\alpha}\,, \label{Eq38}\\
   a_2^{(L)}&=& \frac{1 + \alpha^2}{2\alpha}(A_2 + A_3) + \frac{1 + 3\alpha^2}{\alpha(3 + \alpha^2)}A_3 \,, \nonumber\\
  a_3^{(L)} &=& \frac{1 + 3\alpha^2}{\alpha (3 + \alpha^2)}A_3\,.\nonumber
\end{eqnarray}
Now we consider the solute concentration  in the solid  phase. In
line with the above  we take  the solute concentration $c_S(x) =
C_S|_{X = 0}$ at the interface as
\begin{equation}\label{Eq39}
 c_S (x)  =  a_0^{(S)} + a_1^{(S)}e^{- \gamma_1 x/2} + a_2^{(S)}e^{- \gamma_2 x/2} + a_3^{(S)}e^{- \gamma_3
 x/2}\,.
\end{equation}
 Taking into account that
\begin{eqnarray}\label{Eq40}
J(x, t)|_{X = 0} &=& A_0 - (A_0 + A_2 + A_3)e^{- \gamma_1x/2}\nonumber\\
&&+ A_2 e^{- \gamma_2x/2} +  A_3e^{- \gamma_3x/2}\,,
\end{eqnarray}
one substitutes  the expression  for $c_S(x)$, $c_L(x)$ and $J(x,
t)|_{X = 0} $ in the interface boundary condition (\ref{Eq12}). Then
using the equalities (\ref{Eq39}), (\ref{Eq36}) and (\ref{Eq38}) one
obtains for  the coefficients $a_n^{S}$
\begin{eqnarray}
  a_0^{(S)} &\equiv & c_S^{\infty} = c_0\,,\nonumber\\
  a_1^{(S)} &=& - \frac{2A_0}{\alpha} - \frac{1 - \alpha^2}{2\alpha}(A_2 + A_3)\,,\label{Eq41}\\
  a_2^{(S)} &=&   \frac{3 + \alpha^2 }{2\alpha}(A_2 + A_3) - \frac{2(1 - \alpha^2)}{\alpha (3 + \alpha^2)}A_3\,,\nonumber\\
  a_3^{(S)} &=&  \frac{4(1 + \alpha^2)}{\alpha(3 + \alpha^2)}A_3 \,.\nonumber
\end{eqnarray}
As it is seen from Eqs.~(\ref{Eq35}) and (\ref{Eq37})-(\ref{Eq41})
all the quantities  of interest to us are defined   in terms  of the
parameters $A_0$, $A_2$ and $A_3$. The value of the flux
$J_L^{\infty} = A_0$  at the interface infinitely removed  from
surface can be found  from boundary condition (\ref{Eq12}) taken at
$x \rightarrow \infty $
\[J_L^\infty/\alpha = c_S^\infty - c_L^\infty = c_0 (1 - 1/k_\infty)\,, \]
where $k_\infty(\alpha) = c_S^\infty /c_L^\infty$ is the partition
coefficient for the infinite system given by (\ref{Eq2})
\[
 k_\infty(\alpha) = \left\{\begin{array}{ll}
\displaystyle{\frac{(1 - \alpha^2)k_e + \alpha (V_D/V_{DI})}{1 - \alpha^2 + \alpha (V_D/V_{DI})}}\,,&   \alpha < 1\\
1\,,& \alpha \geqslant 1\,.
\end{array}\right.\]
The constants $A_2$ and $A_3$ can be defined from the boundary
conditions
\[C_S(0) = c_1\hspace{1cm}\frac{\partial C_S}{\partial x}\Bigr |_{x =0} = 0\,,
\]where the second equality means  the condition
of absence of the solute flux from surface of the solid usual for
the Fick's diffusion.

 As regards the inhomogeneous partition
coefficient $k_\alpha(x)$,  allowing for the Eqs.~(\ref{Eq36})  and
(\ref{Eq39}) it can be presented as
\[k_\alpha(x) = \frac{k_{\infty}(\alpha) + f_S(x,\alpha)/c_L^\infty}{1 + f_L(x,\alpha)/c_L^\infty}\,, \]
where
\begin{eqnarray*}
&& f_i(x,\alpha) = a_1^{(i)}e^{- \gamma_1 x/2} + a_2^{(i)}e^{-
\gamma_2 x/2} + a_3^{(i)}e^{- \gamma_3
 x/2}\\
 &&(i = L,S)\,.
\end{eqnarray*}
\begin{figure*}
\includegraphics{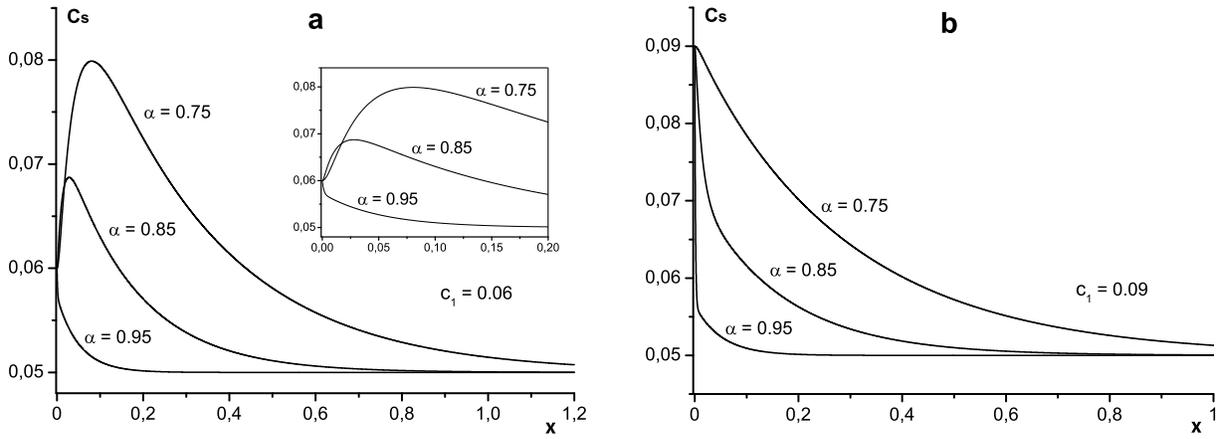}
\caption{\label{fig3} The terminal solute concentration profiles in
the solid; $c_0 = 0.05$, $V_D = 25 (m/s)$, $V_{DI} = 20 (m/s)$, $k_e
= 0.1$.\\ a) $c_1 = 0.06$; b) $c_1 = 0.09$; $\alpha = V/V_D. $ }
\end{figure*}
\begin{figure*}
\includegraphics[-15,25][570,226]{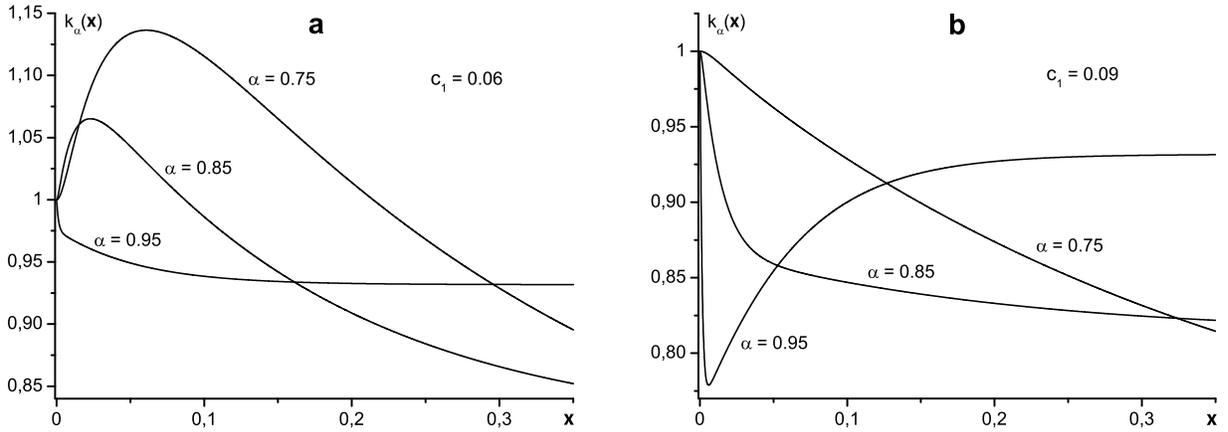}
\caption{\label{fig4} The partition coefficient; $c_0 = 0.05$, $V_D
= 25 (m/s)$, $V_{DI} = 20 (m/s)$, $k_e = 0.1$. a) $c_1 = 0.06$; b)
$c_1 = 0.09$; $\alpha = V/V_D. $ }
\end{figure*}
\begin{figure*}
\includegraphics[10,30][570,236]{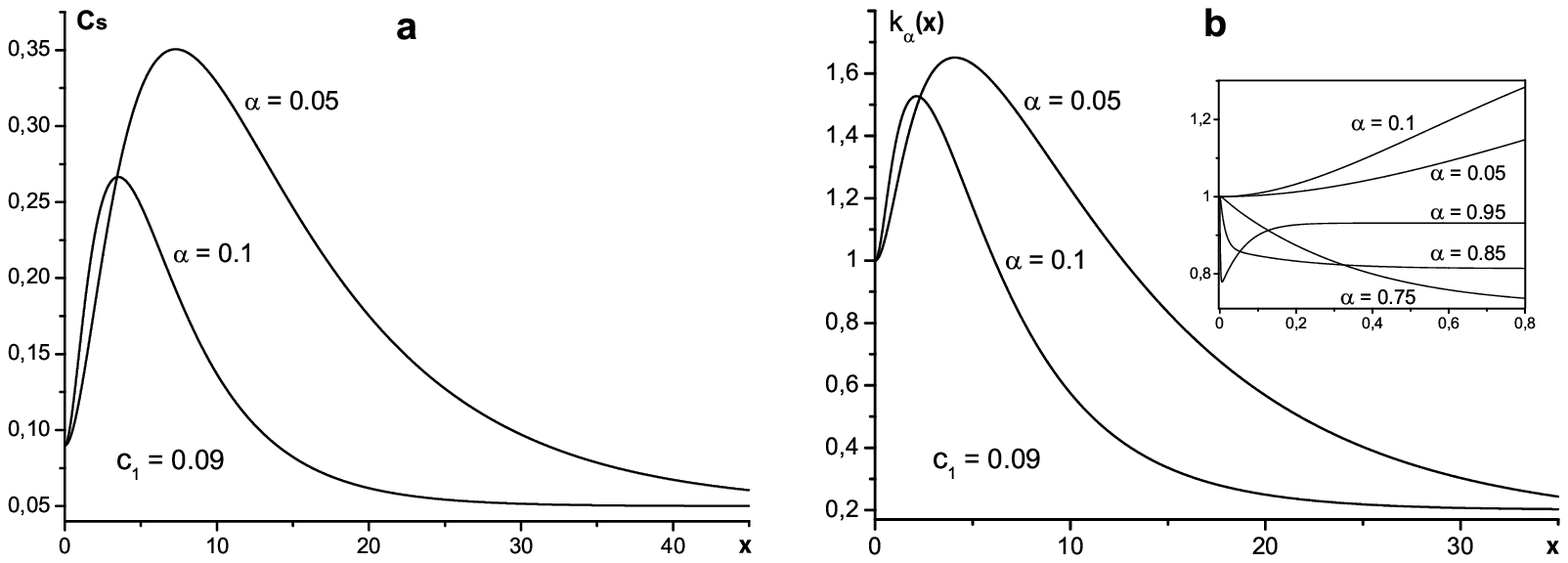}
\caption{\label{fig5} a) The terminal solute concentration profiles
in the solid for the slow motion of the interface. b) the partition
coefficient for the slow motion of the interface. In both case $V_D
= 25 (m/s)$, $V_{DI} = 20 (m/s)$, $k_e = 0.1$, $c_0 = 0.05$; $\alpha
= V/V_D $ .}
\end{figure*}
\section{\label{sec:dis}Discussion}
Figs.~\ref{fig1}--\ref{fig3} present the solute distributions for
$c_0 = 0.05 $ and different  values  of $\alpha = V/V_D$ and $c_1$
for some moments of time $t < 0$. The solid lines are solute
distributions in the solid phase, the dash lines  are the solute
distributions in the liquid. The jump of the concentration takes
place at the point of occurring of the interface. The terminal
solute distributions $C_S(x)$ at $t = 0$ in the solid are given in
Fig.~\ref{fig3}. The behavior of the partition coefficient for
different parameters is shown in Fig.~\ref{fig4}.

As it is seen from Figs.~\ref{fig1}--\ref{fig3}   the concentration
curves are very sensitive to the variation of the parameters
$\alpha$ and $c_1$.  In Fig.~\ref{fig1}a the concentration curves
for $\alpha = 0.95$ and $c_1 = 0.06$ are shown. When the interface
is far enough from the surface the solute distribution has the form
typical for the stationary conditions of a solidification (inset in
Fig.1a). The motion of the interface near the surface gives rise to
the increase of the solute concentration in the liquid phase. The
curve at $t = 0$ defines the terminal solute distribution in the
solid (see also Fig.~\ref{fig3}).

The partition coefficient  for $\alpha = 0.95$ and $c_1 =0.06$ is
shown in Fig.~\ref{fig4}a and exhibits the monotonic increase
reaching   unity at the interface.

At $\alpha = 0.85$ and $c_1 =0.06$ the solute distribution in the
solid behaves not monotonically, having the maximum removed from the
surface (Fig.~\ref{fig1}b). In this case  the partition coefficient
exceeds unity in the near surface region (Fig.~\ref{fig4}a). The
reason may be in the following.

At the high  growth velocity $V \rightarrow V_D$ $(\alpha
\rightarrow 1)$ the interface reaches the surface practically
simultaneously with the concentration disturbance  moving with the
velocity $V_D$.  In this case  the effects of the reflection have no
time  to affect  markedly the value of the solute concentration at
the interface. When the interface  moves not enough rapidly   (for
example, at $\alpha = 0.85$) the concentration wave have time to
reflect at the surface and to reach   the solid phase in the near
surface region. In this case  the resulting solute flux at the
liquid side of the interface will be defined by the sum of two
contributions: the flux  of the solute atoms  rejected by the
interface and the oppositely directed flux of the the atoms
reflected by the surface. In the end competition between them can
lead to a decrease of the resulting flux and at  a later time to its
complete disappearance.

In the inset in Fig.~\ref{fig1}b the solute flux in the liquid phase
is shown. At the interface occurring at the sufficient distance from
the surface the flux is directed to the liquid phase $(t = - 0.14, x
\thickapprox 0.12, J < 0)$. As the interface advances the flux at
the interface practically disappears  (at $t \thickapprox 0.1$). The
flux produced by the interface and the reflected flux will be equal
in  value. The  jump of the concentration  at this moment is absent
and the partition coefficient reaches unity (Fig.~\ref{fig4}a). At a
later time owing to the reflected particle  the resulting flux at
the interface proves to be pointed toward the growth phase playing
the role  of the "external force" increasing the migration of the
solute atoms through  the interface to the solid (the inset in Fig.
1b, $t = - 0.09, J >0$). As a result, this gives rise to rapid
enough growth of the solute concentration at the solid side of the
interface and the decrease at the liquid side of one
(Fig.~\ref{fig1}b,  t= - 0.04, $J
> 0$). Beginning from the moment of  time  $t \approx - 0.1$, the
partition coefficient becomes greater then unity (Fig.~\ref{fig4}a),
reaching
 unity only at the surface.

At Fig.~\ref{fig3} the terminal  solute distributions in the solid
at different values  $\alpha$ and $c_1$  are given.  The
concentration maximums are shifted  to the surface with increasing
$c_1$ (Fig.~\ref{fig3}b, $c_1 = 0.09$). At the same time at $\alpha
= 0.95$ the partition coefficient does not behave monotonically,
reaching  the minimum near the surface. The following circumstance
can be responsible for such behavior.

When the interface moves with the velocity $V \sim V_D$ the
concentration disturbance has no time to propagate to considerable
distances from the source of the disturbance. As a result, the
deviations from the initial concentration $c_0$ will be only
exhibited within the thin  liquid layer near the surface. The
greater is the velocity, the thinner is this layer.  The large
enough value of $c_1$ (relative to $c_0$) presumes the drastic
growth of the solute concentration in such a thin layer in
comparison with the solid side of the interface. In Fig.~\ref{fig2}
it is seen that when the interface moves in the near surface region
the concentration jump increases leading to the decrease  of the
partition coefficient (Fig.~\ref{fig4}b). At the smaller values  of
$c_1$ ($c_1 = 0.06$, $\alpha = 0.95$) the interface advance is
accompanied by the decrease  of the jump (Fig. 1a) and the monotonic
behavior of $k_\alpha$ (Fig.~\ref{fig4}a). The same monotonic
behavior of $k_\alpha$ is appeared at $c_1 = 0.09$ but the smaller
$\alpha$.

\section{\label{sec:con}Conclusion}
In the given work the one dimensional model of the solute  diffusion
during  rapid solidification of the binary alloy in the
semi-infinite volume is presented. Within the scope  of the model it
is supposed  that the planar interface  moves with a constant
velocity, local equilibrium near the interface  in the bulk of the
liquid phase is absent and  the diffusion flux is related to the
concentration gradient  by the Maxwell-Cattaneo equation. The latter
gives rise  to the telegrapher equation for both  the flux and the
solute concentration. The solution of this equation  in the region
occupied  by the melt has been found by the Riemann method
\cite{TS04}. The basic assumption concerns  to  the behavior of the
flux  and the solute concentrations at the  moving phase interface.
Based on  the model form of the partition coefficient given in the
work \cite{AA99} it has been suggested that these quantities should
be given in the form of the expansions (\ref{Eq21}),
 (\ref{Eq22}) and (\ref{Eq36})  (\ref{Eq39}). Such  choice  has made
 it possible  to obtain in an explicit form the expressions for the flux
 and the solute concentrations in the both phases and  to analyze  different regimes
 of rapid solidification in the near surface region.

Within the scope of the model it has been shown that different
 types  of the concentration profiles can be realized depending on
 the external parameters (the growth velocity $V$ and the solute
  concentration at the surface of the solid
 $c_1$).  As regards the partition coefficient, it may  behavior not
 monotonically  reaching both the maximum and the minimum in the
 near surface region.
\begin{figure*}
\includegraphics{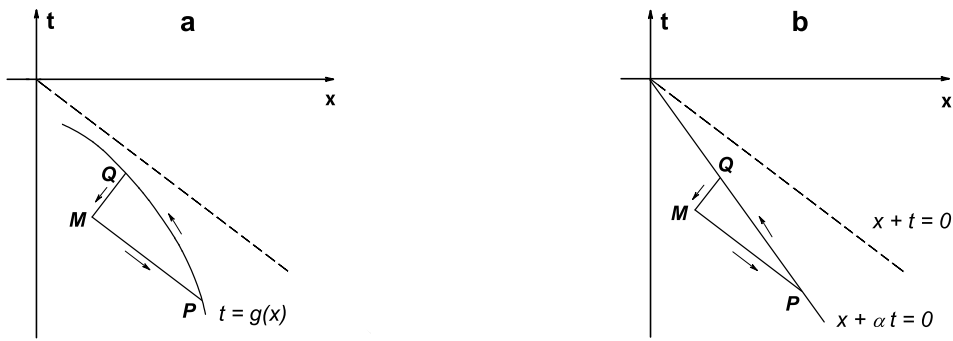}
\caption{\label{fig6}The figures to the Riemann method. }
\end{figure*}

 It should be noted that if the expansions (\ref{Eq21}),
 (\ref{Eq22}) contain the infinite sum of terms, then the expressions
 for the flux  (\ref{Eq35}) and  the solute concentrations  (\ref{Eq37}) and
 (\ref{Eq39}) valid only for $\alpha = V/V_D$ close to unity.  The
 equalities (\ref{Eq32}) defining the number of the terms in the sum
 (\ref{Eq21}) and (\ref{Eq22}) ( similar equalities hold for the expansion (\ref{Eq36})
 as well) have been written by recognizing that at $\alpha \rightarrow
 1$ the terms with $n \geqslant 4$ give the exponential small
 contributions to Eqs.~(\ref{Eq33}), (\ref{Eq35}) and (\ref{Eq37}),
 (\ref{Eq39}). However,  one can consider  the models of
 solidification when the sums  (\ref{Eq21}) and (\ref{Eq22}) contain a
 finite number of terms. In particular, if one  defines  $\varphi (x)$
 from the outset in the form
\begin{equation}\label{Eq44}
\varphi(x) = A_0 + A_1e^{- \gamma_1 x/2} + A_2e^{- \gamma_2 x/2} +
A_3e^{- \gamma_3 x/2}
 \end{equation}
and similarly for $\psi (x)$ and other quantities  of interest, then
when the first condition from the equalities (\ref{Eq32}) are held
the  expressions found for the flux (\ref{Eq33}), (\ref{Eq35}) and
the solute concentrations (\ref{Eq37}), (\ref{Eq39}) will give the
exact solution for this model valid for any  $\alpha < 1$ including
the case of $\alpha \rightarrow 0 $ that corresponds to local
equilibrium and  the Fick's diffusion. In the latter case  it is
easy to verify  that  the expressions (\ref{Eq35}) and  (\ref{Eq37})
satisfy  the diffusion equations.

 The terminal solute distributions
in the solid  for the model (\ref{Eq44}) are given  in
Fig.~\ref{fig5}a for different values $\alpha$ and $c_1 = 0.09$.
Comparing Fig.~\ref{fig5}a with Fig.~\ref{fig3}b it is seen that the
solute distributions at the small $\alpha$ differ rather drastically
from that  at the large $\alpha$.  At the slow interface motion  the
concentration profiles
 does not behave monotonically at the given $c_1$.  The concentration
maximums  are  several times  bigger than  the solute concentration
at the solid surface. The behavior of the partition coefficient  is
presented in Fig.~\ref{fig5}b. The non-monotonic behavior  is also
exhibited. In the near  surface region $k_\alpha$ reaches its
maximum exceeding unity.

Another considerable assumption used in the work is that the
interface remains planar during the whole of the solidification
process. The problem of the stability of the planar interface has
been attracting attention  over  several decades
\cite{MS64,TK86,KT90,HTA98,GD04}).  It has been established  that a
planar interface  is stable for both small enough  growth velocities
and high enough ones. More precisely  there are  the critical
velocities  $V_c$ and $V_a$ so that if $V < V_c$ and $V > V_a $ the
 interface is morphological stable against small perturbations
of its form. In the region $ V_c < V < V_a $ the planar interface is
unstable  and degenerates  in different crystal patterns of cellular
or dendritic morphologies. In the work \cite{GD04}  it has been
shown that for the isothermal solidification in the local
nonequilibrium system $V_a < V_D$.

It has been noted  however  that the theoretical study  of the
stability problem is only based on   using  the steady state regime
of the interface motion in a unbounded medium. The influence of the
system surface on the stability  of the planar front requires
further investigation.
\\
\begin{acknowledgments}
The authors thank P. Galenko for  discussion of the work.
\end{acknowledgments}
\appendix
\setcounter{section}{0}
\renewcommand{\thesection}{\Alph{section}}
\setcounter{equation}{0}
\renewcommand{\theequation}{\Alph{equation}}
\renewcommand{\theequation}{\thesection.\arabic{equation}}
\section{The Riemann method}
Let it be required to find the solution of the linear hyperbolic
equation
\begin{equation}\label{EqA1}
\frac{\partial ^2 J}{\partial t^2} + \frac{\partial J}{\partial t} =
\frac{\partial ^2 J}{\partial x^2}\,,
\end{equation}
satisfying  the initial conditions given at the curve
 $\Gamma$ :
$t = g(x)$ (see Fig.~\ref{fig6}a)
\begin{eqnarray*}
J|_{t = g(x)} &=& j_0(x) \\
 \frac{\partial J}{\partial t}\Bigr |_{t = g(x)} &= &j_1(x)\,.
\end{eqnarray*}
The substitution
  $J = e^{- t/2}u$
makes it possible to lead  Eq.~(\ref{EqA1}) to a
more simple form
\begin{equation}\label{EqA3}
\frac{\partial ^2 u}{\partial x^2}  - \frac{\partial^2 u}{\partial
t^2}  + \frac{1}{4}u = 0\,,
\end{equation}
with the initial conditions
\begin{eqnarray}
  u|_{t = g(x)} &=& j_0(x)e^{g(x)/2} \equiv \varphi_1 (x)\label{EqA4} \\
  \frac{\partial u}{\partial t}\Bigr |_{t = g(x)} &=& ({\textstyle\frac{1}{2}}j_0 +j_1)e^{g(x)/2} \equiv \psi_1(x)\,.\label{EqA5}
\end{eqnarray}
The characteristics of Eq.~(\ref{EqA3}) are the straight lines $x
\pm t = \mbox{const}$. According to the Riemann method \cite{TS04}
if the characteristics go through the point $M$ and intersect with
the curve $\Gamma$ at the points $P$ and $Q$ , then the solution of
Eq.~(\ref{EqA3}) at the point $M$ can be represented as
\begin{widetext}
\begin{equation}
  u(M) = \frac{1}{2}\bigr (u_P + u_Q  \bigl) -  
   \frac{1}{2}\int\limits_{PQ}v\Bigl ( \frac{\partial u}{\partial x_1 }dt_1 + \frac{\partial u}{\partial t_1} dx_1 \Bigr
   ) - u\Bigl ( \frac{\partial v}{\partial x_1 }dt_1 + \frac{\partial v}{\partial t_1} dx_1 \Bigr
   )\label{EqA6}
\end{equation}
\end{widetext}
The integral in (\ref{EqA6}) is taken along the curve  $\Gamma$ from
 $P$ up $Q$ and $u_P$ and $u_Q$ are the values of  $u$, taken at the points  $P$
 and $Q$. The Riemann function $v(M, M_1)$ for Eq.~(\ref{EqA3}) has the  form
\begin{equation}\label{EqA7}
    v(M, M_1) = J_0\Bigl ({\textstyle\frac{1}{2}}\sqrt{(x - x_1)^2 - (t -
    t_1)^2}
    \Bigr )\,,
\end{equation}
where $J_0(x)$  is the Bessel function of zero order and ${\partial
u}/{\partial x}$ is calculated along the curve as
\begin{equation}\label{EqA8}
    \frac{\partial u}{\partial x}\Bigr |_{t = g(x)} = \varphi
    _1'(x) - \psi_1(x)g'(x)\,.
\end{equation}
The Riemann method for arbitrary linear hyperbolic equations one can
find, for example, in \cite{TS04}.

Now consider the solution of  Eq.~(\ref{EqA3}) in the region  $x
\geq 0$, $t \leq 0$, $X = x + \alpha t < 0 $, when  the initial data
are given at the straight line  $ t = - x/\alpha$ (see
Fig.~\ref{fig6}a). Instead of (\ref{EqA4}) and (\ref{EqA5}) we have
\begin{eqnarray}
 u|_{t = - x/\alpha} &=& j_0(x)e^{-x/2\alpha} \label{EqA9} \\
  \frac{\partial u}{\partial t}\Bigr |_{t = - x/\alpha} &=& ({\textstyle\frac{1}{2}}j_0 +j_1)e^{-x/2\alpha} \,.\label{EqA10}
\end{eqnarray}
If the point $M$ has coordinates  $(x, t)$, so it is easy  to show
that the points $P$ and $Q$ have the abscissas respectively equal to
\begin{equation}\label{EqA11}
    x_P = - \frac{\alpha (x + t)}{1 - \alpha}\, ;\hspace{1 cm} x_Q =  \frac{\alpha (x - t)}{1  + \alpha }
\end{equation}
Consider the integral term in Eq.~(\ref{EqA6}). Using
Eqs.~(\ref{EqA8})-(\ref{EqA11}) and the fact that along  the pathway
of integration $dt_1 = - dx_1/\alpha$, one has
\begin{widetext}
\begin{equation}\label{EqA12}
\frac{1}{2}\int\limits_{\displaystyle{-\frac{\alpha (x + t)}{1 -
\alpha}}}^{\displaystyle{\frac{\alpha (x - t)}{1 + \alpha}}}dx_1
e^{- x_1/2\alpha}\Bigl \{v\psi (x_1) + \varphi  (x_1) \Bigl (
\frac{1}{\alpha }\frac{\partial v}{\partial x_1} - \frac{\partial
v}{\partial t_1} \Bigr ) \Bigr\}_{t_1 =- x_1/\alpha}\,,
\end{equation}
\end{widetext}
where the notations are introduced
\begin{eqnarray}
  \varphi (x) &=& j_0(x)\,, \nonumber \\
  \psi (x) &=& \frac{1}{2}j_0(x) - \frac{1}{\alpha }j_0'(x) - \frac{1 - \alpha ^2}{\alpha
  ^2}j_1(x)\,.\nonumber
\end{eqnarray}
Furthermore using  the Riemann function (\ref{EqA7}), it can show
that
\begin{eqnarray}\label{EqA13}
\Bigl (\frac{1}{\alpha }\frac{\partial v}{\partial x_1} -
\frac{\partial v}{\partial t_1}\Bigr )\Bigl |_{t_1 =
 - x_1/\alpha} = \phantom{aaaaaaaaaaaaa} &&\nonumber\\
 - \frac{X}{2\alpha} \frac{J_0' \Bigl
(\frac{1}{2}\sqrt{(x - x_1)^2 - (t + x_1/\alpha)^2} \Bigr
  )}{\sqrt{(x - x_1)^2 - (t + x_1/\alpha)^2}}. &&
\end{eqnarray}
\\
At last, after of the substitution the integral (\ref{EqA12}) into
Eq.~(\ref{EqA6}) and using  the equality $J = e^{- t/2}u$, one
obtains the solution of the starting equation (\ref{EqA1}), with
added conditions (\ref{Eq16}) and (\ref{Eq17}), in the form
represented by Eq.~(\ref{Eq18}) in Sec. \ref{sec:mod}.
\\

\appendix
\setcounter{section}{1}
\renewcommand{\thesection}{\Alph{section}}
\setcounter{equation}{0}
\renewcommand{\theequation}{\Alph{equation}}
\renewcommand{\theequation}{\thesection.\arabic{equation}}
\section{The calculation of the integrals}

Substituting Eq.~(\ref{Eq21}) and (\ref{Eq22}) into (\ref{Eq18}) we
have
\begin{equation}\label{EqB1}
J(x, t) = \sum\limits_{n\geqslant 0}J_n(x, t)\,,
\end{equation}
where
\begin{widetext}
\begin{equation}
 J_n(x, t) =
-  B_nJ_n^{(1)} + A_nJ_n^{(2)}
 +\frac{A_n}{2}\Bigl\{\exp\Bigl[\frac{\alpha\gamma_n(x + t) +
X}{2(1 - \alpha)}\Bigr ]
 + \exp\Bigl[- \frac{\alpha\gamma_n(x - t) + X}{2(1 +
\alpha)}\Bigr ]  \Bigr\}\,;\label{EqB2}
\end{equation}
and
\begin{eqnarray}
   J_n^{(1)}&=&  \frac{1}{2}e^{- t/2}\int\limits_{\displaystyle{-\frac{\alpha (x + t)}{1 - \alpha}}}^{\displaystyle{\frac{\alpha (x - t)}{1 + \alpha}}}
  \,dx_1 e^{- x_1/2\delta_n}J_0 \Bigl (\frac{1}{2}\sqrt{(x - x_1)^2 - (t + x_1/\alpha)^2}  \Bigr )\,;\label{EqB3}\\
  J_n^{(2)}&=&\frac{X}{4\alpha}e^{- t/2}\int\limits_{\displaystyle{-\frac{\alpha (x + t)}{1 - \alpha}}}^{\displaystyle{\frac{\alpha (x - t)}{1 + \alpha}}}
  \,dx_1 e^{- x_1/2\delta_n}\frac{J_0' \Bigl (\frac{1}{2}\sqrt{(x - x_1)^2 - (t + x_1/\alpha)^2}  \Bigr
  )}{\sqrt{(x - x_1)^2 - (t + x_1/\alpha)^2}}\,;\label{EqB4}\\
    \delta_n& =& \frac{\alpha}{1 + \alpha\gamma_n}\label{EqB5}
\end{eqnarray}
\end{widetext}

\subsection*{The calculation  $J^{(1)}_n$}
Making the substitution in the integral (\ref{EqB3})
\[\frac{2\alpha X}{1 - \alpha ^2}z = \frac{\alpha (x + t)}{1 - \alpha} +
x_1\,,\] we have (for convenience the index $n$ is omitted)
\begin{equation}\label{EqB6}
    J^{(1)} = \frac{\alpha X}{1 - \alpha ^2}\exp \Bigl [ \frac{X'}{2(1 - \alpha
    )}\Bigl ]\mathscr J\,,
\end{equation}
where the notations are introduced
\begin{eqnarray}
  &{\mathscr J} &= \int\limits ^1_0 e^{- \mu z} J_0\Bigl (\beta \sqrt{z(1 - z)}\Bigr )dz \,,\label{EqB7} \\
  &X'&= X + \Bigl (\frac{\alpha}{\delta} - 1 \Bigr )(x + t)\label{EqB8}\, ,  \\
 \mu &=& \frac{\alpha X}{\delta (1 - \alpha ^2)} < 0\,,\quad  \beta = - \frac{X}{\sqrt{1 - \alpha ^2}} > 0\,. \label{EqB9}
\end{eqnarray}
Consider the integral ${\mathscr J}$. Using the definition of the
Bessel function
\[ J_0 \Bigl ( \beta\sqrt{z - z^2}\Bigr ) = \sum\limits^\infty _{m = 0}\frac{(-1)^m (\beta /2)^{2m}(z - z^2)^m }{m!\,\Gamma (m + 1)} \,,\]
where $\Gamma (x)$  is the Euler gamma-function, one represents  the
integral (\ref{EqB7}) in the form
\begin{equation}\label{EqB10}
{\mathscr J} = \sum\limits ^\infty _{m = 0}\frac{(-1)^m (\beta
/2)^{2m}}{m!\,\Gamma (m + 1)}\int \limits _0^1 e^{- \mu z}(z -
z^2)^mdz
 \end{equation}
Calculating  the latter integral \cite{GR71}, one obtains
 \begin{equation}\label{EqB11}
{\mathscr J} =  \Bigl ({\pi}/{|\mu|}\Bigr )^{1/2}\; e^{-
\mu}\sum\limits_{n = 0}^{\infty}\frac{(- \beta ^2/4|\mu|)^m}{m!}I_{m
+ 1/2}\Bigl (\frac{|\mu|}{2} \Bigr )\,,
 \end{equation}
where $I_{\nu}(x)$ is the modified Bessel function of the first
kind.  Furthermore, we use the equality \cite{PBM83}
\begin{eqnarray}\label{EqB12}
   &&\sum\limits_{m = 0}^{\infty}\frac{t^m}{m!}I_{m + 1/2}(z) =  \Bigl
    ( \frac{2t}{z} + 1 \Bigr )^{- 1/4}I_{1/2}\Bigl (\sqrt {z^2 + 2tz }   \Bigr
    )\nonumber\\
    &&|z| - |2t| > 0\,.
\end{eqnarray}
In our case
\[|z| - |2t| = \frac{\delta |X|}{2\alpha(1 - \alpha ^2)}(\alpha ^2\gamma ^2 + 2 \alpha \gamma + \alpha ^2) > 0
\] and instead of Eq.~(\ref{EqB11}) we have
\begin{equation}\label{EqB13}
{\mathscr J} = \sqrt{\frac{\pi}{\nu|\mu|}}\; e^{- \mu/2}I_{1/2}\Bigl
(\frac{\nu|\mu|}{2} \Bigr )\,,
\end{equation}
where
\begin{eqnarray}\label{EqB14}
    \nu = \nu (\delta) &=& \sqrt{1 - \frac{\delta ^2}{\alpha ^2}(1 -
    \alpha ^2)}\nonumber\\
    &=& \frac{\delta }{\alpha }\sqrt{\alpha ^2\gamma ^2 + 2\alpha\gamma
+ \alpha
    ^2}\,.
\end{eqnarray}
Substituting the expression (\ref{EqB13})  into Eq.~(\ref{EqB6}) and
taking into account that $I_{1/2}(x) = (2/\pi x)^{1/2}\sinh (x)$, we
obtain
\begin{widetext}
\begin{equation}\label{EqB15}
    J^{(1)}_n = \frac{\delta _n}{\nu _n} \exp\Bigl [ \frac{X'}{2(1 - \alpha)}\Bigr
    ]\Bigl\{\exp\Bigl [- \frac{\alpha (1 - \nu _n)X}{2\delta _n(1 - \alpha ^2)}{} \Bigr ] -
    \exp\Bigl [- \frac{\alpha (1 + \nu _n)X}{2\delta _n(1 - \alpha ^2)}{} \Bigr ]
    \Bigr\}\,,
\end{equation}
\end{widetext} where $\nu _n = \nu(\delta _n)$. At last,
substituting $X'$ from Eq.~(\ref{EqB8}) into Eq.~(\ref{EqB15}) one
has
\begin{equation}\label{EqB16}
     J^{(1)}_n = \frac{\delta _n}{\nu _n} e^{- \gamma _n x/2}\Bigl\{ \exp\Bigl[\frac{\gamma_n^{(+)}X}{2(1 - \alpha^2)}\Bigr ]
     -
    \exp\Bigl[\frac{\gamma_n^{(-)}X}{2(1 - \alpha^2)}\Bigr ] \Bigr\}
\end{equation}
and \[\gamma_n^{(\pm)} =  \alpha + \gamma_n \pm
\sqrt{\alpha^2\gamma_n^2 + 2\alpha\gamma_n + \alpha^2}\,.\]

\subsection*{The calculated $J_n^{(2)}$}
Consider the integral $J_n^{(2)}$. After substitution of the
variable in Eq.~(\ref{EqB4})
\begin{equation}\label{EqB17}
    \xi + \frac{X}{1 - \alpha ^2} = x - x_1
\end{equation}
we have  (the index $n$ is omitted)
\begin{widetext}
\begin{eqnarray}
J^{(2)}&=& - \frac{X}{4\alpha}\exp{\Bigl [\frac{X'}{2(1 - \alpha)} - \frac{\alpha X}{2\delta (1 - \alpha ^2)} \Bigr ]}\times \nonumber\\
  & &\hspace{2cm}\times\int\limits_{\displaystyle \frac{\alpha X}{1 - \alpha ^2}}^{\displaystyle - \frac{\alpha X}{1 - \alpha
    ^2}}d\xi   e^{\xi/2\delta } \frac{J'_0\Biggl (\frac{1}{2}
    \sqrt{{\displaystyle\frac{1 - \alpha ^2}{\alpha ^2}}}\sqrt{\Bigl ({\displaystyle\frac{\alpha X}{1 - \alpha ^2}\Bigr )^2} - \xi ^2} \; \Biggr )}
    {\sqrt{{\displaystyle\frac{1 - \alpha ^2}{\alpha ^2}}}\sqrt{\Bigl ({\displaystyle\frac{\alpha X}{1 - \alpha ^2}\Bigr )^2} - \xi^2}
    }\,.\label{EqB18}
\end{eqnarray}
To calculate  the integral (\ref{EqB18}) we consider the equality
(\ref{EqB15}), having previously  made the substitution
(\ref{EqB17}) into $J_n^{(1)}$. After reducing common factors, we
have
\begin{equation}
  \int\limits_{\displaystyle \frac{\alpha X}{1 - \alpha ^2}}^{\displaystyle - \frac{\alpha X}{1 - \alpha
    ^2}}d\xi   e^{\xi/2\delta } J_0\Biggl (\frac{1}{2}
    \sqrt{{\displaystyle\frac{1 - \alpha ^2}{\alpha ^2}}}\sqrt{\Bigl ({\displaystyle\frac{\alpha X}{1 - \alpha ^2}\Bigr )^2} - \xi ^2} \; \Biggr )
 = - \frac{4\delta}{\nu}   \sinh \Bigl [\frac{\alpha\nu X}{2\delta(1 - \alpha ^2)}  \Bigr
 ]\,.\label{EqB19}
\end{equation}
Differentiating the latter equation  with respect to $X$, one
obtains
\begin{eqnarray}
    \frac{X}{4\alpha}\int\limits_{\displaystyle \frac{\alpha X}{1 - \alpha ^2}}^{\displaystyle - \frac{\alpha X}{1 - \alpha
    ^2}}d\xi   e^{\xi/2\delta } \frac{J'_0\Biggl (\frac{1}{2}
  \sqrt{{\displaystyle\frac{1 - \alpha ^2}{\alpha ^2}}}\sqrt{\Bigl ({\displaystyle\frac{\alpha X}{1 - \alpha ^2}\Bigr )^2} - \xi ^2} \; \Biggr )}
    {\sqrt{{\displaystyle\frac{1 - \alpha ^2}{\alpha ^2}}}\sqrt{\Bigl ({\displaystyle\frac{\alpha X}{1 - \alpha ^2}\Bigr )^2} - \xi^2}
    } &=&  \nonumber\\
   = \cosh\frac{\alpha X}{2\delta (1 - \alpha ^2)} &-& \cosh\frac{\alpha\nu X}{2\delta (1 - \alpha ^2)}\,.\label{EqB20}
\end{eqnarray}
One multiplies the  latter equality  by
\[ - \exp\Bigl[ \frac{X'}{2(1 - \alpha)} - \frac{\alpha X}{2\delta (1 - \alpha ^2)} \Bigr] \]
and using Eqs.~(\ref{EqB18}), (\ref{EqB5}) and (\ref{EqB8}), one has
\begin{eqnarray}
  J_n^{(2)} &=& \frac{1}{2}\,e^{- \gamma_nx/2}\Bigl\{\exp\frac{\gamma_n^{(+)}X}{2(1 - \alpha ^2)} +
  \exp\frac{\gamma_n^{(-)}X}{2(1 - \alpha ^2)} \Bigr\} - \nonumber\\
  & & \hspace{1cm}- \frac{1}{2}\Bigl\{ \exp\Bigl [\frac{\alpha\gamma _n(x + t) + X}{2(1 - \alpha)}\Bigr ] +
   \exp\Bigl [- \frac{\alpha\gamma _n(x - t) + X}{2(1 + \alpha)}\Bigr ] \Bigr\}\,.\label{EqB21}
\end{eqnarray}
At last, substitute Eqs.~(\ref{EqB16}) and (\ref{EqB21}) into
Eq.~(\ref{EqB2})  and as a result we have
\newpage
\begin{equation}\label{EqB22}
    J_n(x t) = e^{- \gamma_n x/2}\Bigl\{ A_n^{(-)}\exp\Bigl[\frac{\gamma_n^{(+)}X}{2(1 - \alpha^2)}\Bigr ] +
    A_n^{(+)}\exp\Bigl[\frac{\gamma_n^{(-)}X}{2(1 - \alpha^2)}\Bigr ]
    \Bigr\}\,,
\end{equation}
where
 \[ A_n^{(\pm)}= \frac{A_n}{2} \pm
 B_n\frac{\delta_n}{\nu_n}\,. \]
\end{widetext}


\begin{thebibliography} {99}
\bibitem{COM}
The equation $\gamma^{(+)}_n = \gamma^{(+)}_{n - 1 }$ either has no
the solutions or does not give the new ones.
\end{thebibliography}
\end{document}